\documentclass[letter]{aa}
\usepackage{graphicx}
\usepackage{txfonts}
\begin{document}
   \title{Globular clusters and dwarf galaxies in Fornax}
   \subtitle{I. Kinematics in the cluster core from multi-object 
spectroscopy\thanks{Based on observations collected at the ESO La Silla and 
Paranal observatories (programs 66.A--0345, 072.A--0389 and 
074.A--0756).}\fnmsep\thanks{Tables \ref{gc}--\ref{dw} are only available in electronic 
form at the CDS via anonymous ftp to {\tt cdsarc.u-strasbg.fr (130.79.128.5)} 
or via {\tt http://cdsweb.u-strasbg.fr/cgi-bin/qcat?J/A+A/vvv/ppp}}
}
\author{G.~Bergond\inst{1,2,3}
  \and E.~Athanassoula\inst{4}
  \and S.~Leon\inst{5}
  \and C.~Balkowski\inst{2}
  \and V.~Cayatte\inst{6}
  \and L.~Chemin\inst{2}
  \and R.~Guzm\'an\inst{7}
  \and G.~Meylan\inst{8}
  \and Ph.~Prugniel\inst{2,9}}
\institute{Instituto de Astrof\'isica de Andaluc\'\i a,
  C/ Camino Bajo de Hu\'etor 50, 18008 Granada, Espa\~na; 
  \email{gbergond@caha.es}
  \and  Observatoire de Paris, GEPI (CNRS UMR 8111 \& 
Universit\'e Paris 7), 5 place Jules Janssen, 92195 Meudon, France 
\and Department of Physics and Astronomy, Michigan State University,
  East Lansing,  MI 48824, USA
  \and Observatoire de Marseille, 2 place Le Verrier, 13248 Marseille 
Cedex 04, France 
  \and Instituto de Radioastronom\'\i a Milim\'etrica (IRAM), Avda. 
Divina Pastora 7, local 20, 18012 Granada, Espa\~na
\and Observatoire de Paris, LUTH (CNRS UMR 8102 \& 
Universit\'e Paris 7), 5 place Jules Janssen, 92195 Meudon, France  
  \and Astronomy department, Univ. of Florida, 211 Bryant Space 
Science Center, P.O. Box 112055, Gainesville, FL 32611, USA
  \and \'Ecole Polytechnique F\'ed\'erale de Lausanne (EPFL), Laboratoire 
d'Astrophysique, Observatoire, 1290 Sauverny, Suisse 
  \and Centre de Recherches Astronomiques, Universit\'e Lyon 1, 
Observatoire de Lyon, 69561 Saint Genis Laval Cedex, France
}

   \date{Received XX; accepted XX}

   \abstract
{}
{We acquired radial velocities of a significant number of globular clusters 
(GCs) on wide fields between galaxies in the nearby Fornax cluster of 
galaxies, in order to derive their velocity dispersion radial profile 
and to probe the dynamics of the cluster.
}
{We used FLAMES on the VLT to obtain accurate velocities for 149 GCs, within a 
$\approx$\,500\,$\times$150~kpc 
strip centered on NGC~1399, the Fornax central 
galaxy. These objects are at the very bright tail 
($M_V \la -9.5$) of the 
GC luminosity function, overlapping the so-called ``ultra-compact dwarfs'' 
magnitude range. Eight of the brightest FLAMES-confirmed members indeed show 
hints of resolution in the subarcsecond pre-imaging data we used for selecting 
the $\sim$500 targets for FLAMES spectroscopy.
}
{Ignoring the GCs around galaxies by applying 3$d_{25}$ diameter masks,
we find 61 GCs of 20.0 $\le V \le$ 22.2 lying in the intra-cluster 
(IC) medium. The velocity dispersion of the population of ICGCs is 
200~km\,s$^{-1}$ at $\sim$150~kpc from the central NGC~1399 and rises to 
nearly 400~km\,s$^{-1}$ at 200~kpc, a value which compares with the velocity 
dispersion of the population of dwarf galaxies, thought 
to be infalling from the surroundings of the cluster. 
}
{}
   \titlerunning{Globular clusters and dwarfs in Fornax. I}
   \maketitle
%

\section{Introduction}
Stars between galaxies in groups and in clusters have long been observed  
as a diffuse component -- the intra-cluster (IC) light -- originating 
from unresolved populations. A decade ago, the first Virgo IC objects were 
confirmed as planetary nebulae~(PNe, Arnaboldi et~al. \cite{arnaboldi96}). 
Fornax, as the second nearest cluster of galaxies (20.1\,Mpc, Dunn \& Jerjen 
\cite{dunn06}), is the other ideal dense environment to search for 
different types of intracluster material. 

Globular clusters (GCs) are among the best candidates for IC objects. 
Being compact stellar systems ($r_{\rm{eff}} \la 5$~pc) of 
$-10 \la M_V$, they can be detected by deep imaging as a swarm of 
point-like sources surrounding their host galaxies. Spectroscopic follow-up 
on 8--10 m telescopes makes them unique dynamical tracers in the outer 
halos of nearby galaxies, in particular early-types 
(e.g. Bergond et~al.~\cite{bergond06}). 

West et~al. (\cite{west95}; see also Muzzio \cite{muzzio87}) argued 
that GC populations exist in the IC medium within groups and clusters of 
galaxies. In Fornax, the first ICGCs candidates were proposed by Bassino 
et~al.\  (\cite{bassino03}) in the surroundings of dwarfs, and were observed 
between giant galaxies by Bassino et~al. (\cite{bassino06a}). 
In Virgo, an ICGC population has been detected indirectly by 
wide-field imaging (Tamura et~al. \cite{tamura06}) and four ICGCs 
have been partly resolved into stars in a recent HST/ACS observation
(Williams et~al. \cite{williams06}).
But, to date, no spectroscopically confirmed population of ICGCs has been 
found. Such objects would be ``wandering'' in the cluster potential well, 
probing it as a whole. 

In this {\it Letter}, we present the results of a systematic search for
ICGCs in the nearby Southern Fornax cluster, making use of wide-field
images to select GC candidates and a medium resolution spectroscopic 
follow-up with FLAMES on the VLT.

   \begin{figure*}
\sidecaption
\includegraphics[width=14.5cm]{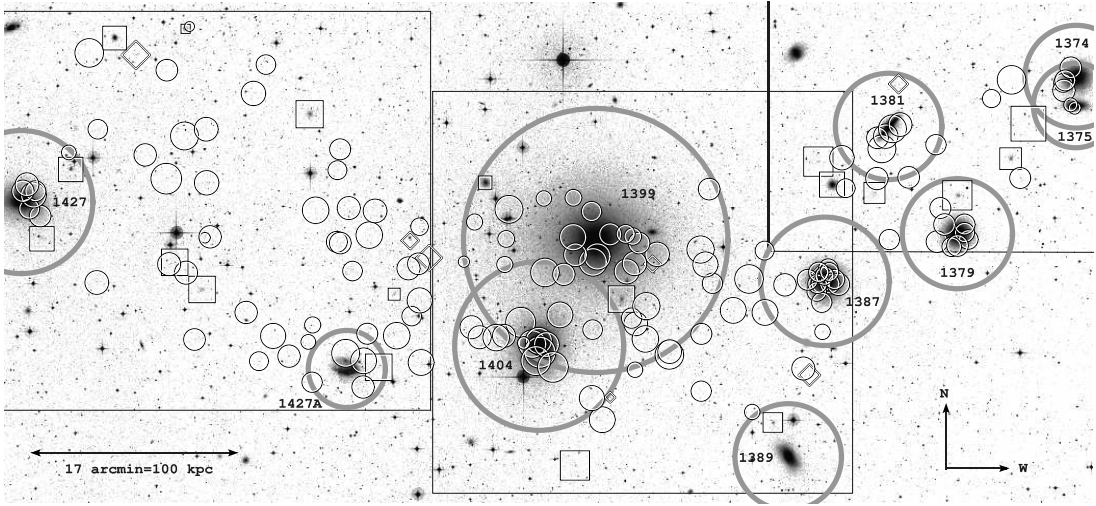}
   \caption{DSS image of the 90$'$$\times$40$'$ ($\sim$530$\times$230~kpc) 
central region of Fornax. The three WFI pre-imaging rectangular (sub-)fields 
are indicated. The large, bold grey circles around each NGC-labelled giant 
Fornax galaxy are the 3$d_{25}$ diameter masks used to select ICGCs. Overlayed 
symbols are the FLAMES confirmed members: circles for the 149 (IC)GCs, squares 
for 19 Fornax dwarfs, and lozenges for 8 ultra-compact objects.
Symbol size is proportional to the velocity, from 548 to 2373~km\,s${^{-1}}$.
}
    \label{map}
    \end{figure*}

\section{The data and inferred GIRAFFE radial velocities}
We give here only a very brief description of the observing runs, data
reduction and analysis and will present more detailed information 
elsewhere. In a separate study (Bergond et~al., in preparation), we obtained 
deep, wide field images of a $\sim$0\fdg9$^2$ strip in Fornax, in $BV(R)I$, 
with the WFI mosaic camera at the La Silla ESO/MPG 2.2-m telescope. 
With its 34$'\times$33$'$ field, WFI is particularly well-suited for
selecting candidate (IC)GCs for spectroscopic follow-up studies. These are
selected as point-like sources having $B-V$ and $V-I$ colours
similar to Galactic GCs and $19.5 \le V \le$ 22.2. 

The spectroscopic follow-up was done using the GIRAFFE/MEDUSA mode 
of FLAMES, the multi-object fibre-fed spectrograph on the UT2 of the VLT.  
We allocated most of the 130 MEDUSA fibres which can be deployed in a
25$'\diameter$ field to $>$100 of the WFI candidate GCs. Our magnitude
selection criteria given above ensure that we get spectra of 
sufficient $S/N$ in $\sim$4-hour exposures with the LR4 grating. 
We completed the target list with slighlty resolved,
$V\le 20.5$ and $V-I<1.5$ objects -- candidate for ultra-compact dwarfs, 
see Sect.~\ref{sectUCDs} -- 
and with visually-selected dwarf galaxies, mostly dE,N. 
Five slightly overlapping FLAMES pointings, observed during two Paris 
Observatory GTO nights, cover a bended strip of $\sim$80$'$$\times$$25'$
around NGC~1399 and include nine other giant Fornax galaxies 
(1$'\simeq$ 5.85~kpc at the adopted distance of 
20.1~Mpc, Fig.~\ref{map}). 
 
A total of more than 500 WFI candidate GCs and dwarfs were followed-up.
Data reduction of all spectra was performed with the IRAF {\sc hydra} package, 
and radial velocities ($\varv$) computed by cross-correlation with the 
{\sc rv}/{\tt fxcor} task, using simultaneously obtained templates and 100 
\'ELODIE library F5--K0 templates. Many Mg/Fe strong absorption lines fall in 
the $\sim$5000--5800~\AA\ wavelength range of the LR4 grating used. According 
to the percentage of valid Gaussian fits to the cross-correlation function 
(CCF) peak, targets were allocated to three 
classes (A, B and C, 
similarly defined in 
Bergond et al. \cite{bergond06}) reflecting the quality of $\varv$ 
estimates.
 
Class A objects represent very secure $\varv$ estimates, where 98\% or more of 
the \'ELODIE templates, as well as more than half of the simultaneous 
templates, agree within the velocity errors. Class B show 60\%--97\% \'ELODIE 
agreement but less than half of the simultaneous templates give consistent 
velocities. Class C have less than 60\% agreeement between \'ELODIE templates; 
these possible members would require a definitive, 
independent confirmation (e.g. observations with other GIRAFFE gratings). 
These are mainly either metal-poor objects showing few lines in the LR4 
grating range, or low signal ($S/N\la5$ per pixel) spectra from the 
faintest sources or from fibres located near the FLAMES field border where 
some vignetting and differential refraction induce flux lost. The  
astrometric accuracy of our input catalogue should not be at stake, as we 
obtained $rms$$<$0\farcs2 w.r.t. the GSC2.1 and UCAC1 catalogues.
  
The typical error $\overline{\delta\varv}$  as estimated by {\tt fxcor} 
is of the order of 10~km\,s${^{-1}}$, and this is confirmed when comparing 
a dozen of objects in common between different FLAMES pointings. Velocities 
for the 15 GCs previously observed in other studies also fairly agree, 
FLAMES reducing the errors by a factor $\sim$2 to $\sim$10 w.r.t. the 
9 GCs in common with Dirsch et~al. (\cite{dirsch04}) and the 6 from Mieske 
et~al. (\cite{mieske04}).

All Fornax members were selected in the velocity range $500 < \varv < 2500$
km\,s$^{-1}$ (Drinkwater et~al. \cite{drinkwater01}), giving a total of
149 (class A+B) GCs and 27 (all classes) dwarfs. In the finding chart 
(Fig.~\ref{map}) and the kinematical analysis hereafter, we do not consider 
class C GCs (but keep class C dwarfs, due to small numbers). 
Our experience with various FLAMES datasets and CCF 
techniques makes us confident that all class A and almost all class B GCs and
dwarfs are {\it bona fide} Fornax members. 
Moreover, excluding class B GCs 
in the following discussion does not change the main results.  
The coordinates and radial velocities of all detected members 
are given in Tables \ref{gc}--\ref{dw}.

\section{Intracluster globular cluster kinematics}
We define as GCs the sources which are unresolved in the WFI study and
whose velocities as found by FLAMES select them as Fornax members 
(Table~\ref{gc}). Note that 
only the bright tail of the GC populations ($M_V \la -9$.5) can be 
observed by FLAMES because of the typical limiting magnitude $V\sim 22$. To 
our knowledge, 134 of these GCs were previously undetected, lying in 
particular at large distances from galaxies. Figure~\ref{map} nevertheless
illustrates the clear concentration of GCs around the Fornax giant 
ellipticals.  To select ICGCs, we mask out all GCs lying close to the ten
giant galaxies in the FLAMES fields. We have taken the mask diameter
to be 3$d_{25}$, the latter obtained from HyperLEDA.

Using these criteria, we find that 61 GCs are part of the intracluster Fornax 
medium. The most striking examples are the GCs situated more than 200~kpc away 
-- in projection -- from NGC~1399; some of these ``wanderer'' GCs being 
located 100~kpc from any bright galaxy. Of course, by masking 1.5$d_{25}$  
around each of the galaxies we miss true ICGCs which are projected close 
to Fornax giants. The masked 
areas represent around a third of the total FLAMES coverage, 
so about 20 ICGCs should be occulted (considering they 
have an uniform spatial distribution). Compared to the 88 GCs we count within 
the 3$d_{25}$ of all ten galaxies, the masked area should therefore include 
around 23\% contamination by projected ICGCs.

\begin{figure}
{\includegraphics[width=8.6cm]{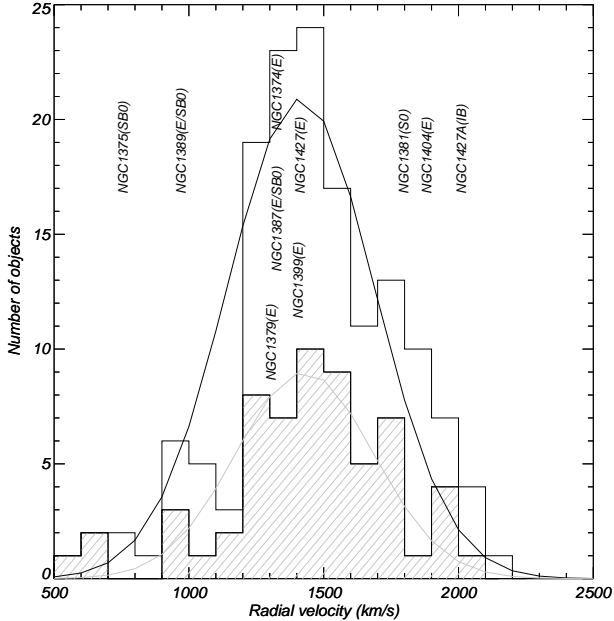}}
   \caption{Histograms of radial velocities for all 149 GCs in Fornax 
(solid line) and the subsample of 61 ICGCs (hatched area). Names of the main 
Fornax galaxies in our fields are located in abcissa at their systemic 
velocities, with the galaxy morphology following between parenthesis.
}
    \label{hist}
    \end{figure}

The histogram of the GC velocities (Fig.~\ref{hist}) shows that the velocity
distribution of the ICGCs is more symmetric than that of the total GC 
population. The latter seems skewed by GCs with velocities around those of 
NGC~1404 and NGC~1427A, i.e. high w.r.t. that of the cluster. Fitting a 
Gaussian to each of the distributions, we find that the velocity peaks are 
similarly located at 1415$\pm$7 
(resp. 1429$\pm$8)~km\,s$^{-1}$ for all (resp. IC) GCs, 
with a $FWHM$ of 644 
(resp. 602) 
km\,s${^{-1}}$.  The ICGC global projected velocity dispersion is
$\overline{\sigma_{\rm p}} = 312^{+8}_{-5}$~km\,s${^{-1}}$. 

\begin{figure}   
{\includegraphics[width=8.6cm]{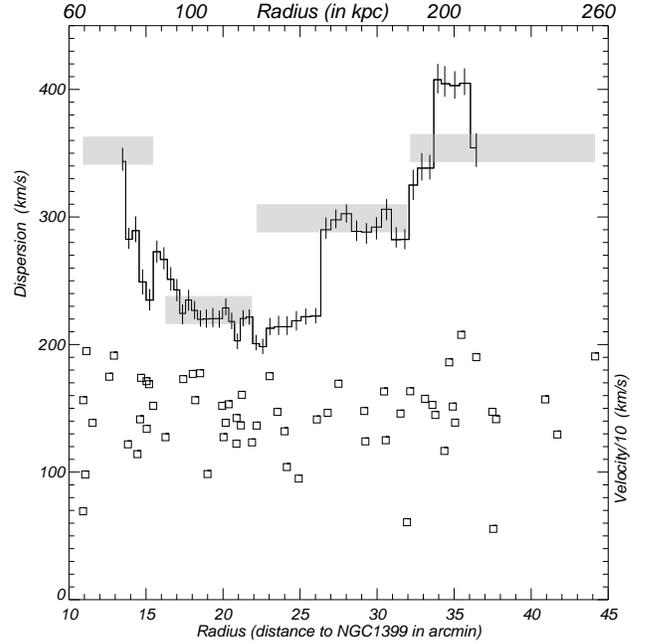}}
   \caption{Velocity dispersion $\sigma_{\rm p}$ radial profile for ICGCs 
in Fornax as a function of radius $R$ (projected distance to NGC~1399), 
using a sliding window containing 20 ICGCs (black curve with error
bars) as well as $\sigma_{\rm p}$ computed in 4 independent bins containing
15 ICGCs (16 for the outermost bin), shown in light grey; the 
height corresponds to the error on $\sigma_{\rm p}$.
The dip at $\sim$150~kpc might correspond to the transition from the baryonic 
mass profile of the central galaxy to its extended dark matter halo. Below the 
$\sigma_{\rm p}$ curve are shown the $\varv/10$ velocities of the 61 ICGCs. 
}
    \label{disp}
    \end{figure}

Figure~\ref{disp} shows the $\sigma_{\rm p}$ of ICGCs as a function of the 
projected distance to NGC~1399, $R$, as estimated within a sliding window  
including 20 GCs. This choice gives in each thus defined radial bin error 
bars $\la$\,20~km\,s$^{-1}$ (as estimated from bootstrap uncertainties).
Overlayed (light grey bars) are the $\sigma_{\rm p}$ values computed in four 
independant radial bins of 15 GCs each (16 for the outermost one). 
Both estimates agree. 

From 75 to 100~kpc a sharp decrease of  $\sigma_{\rm p}(R)$ is 
found,  with a strong gradient of about $-$5~km\,s${^{-1}}$\,kpc$^{-1}$. 
Between 100 and 150~kpc, $\sigma_{\rm p}(R)$ remains rather constant around 
200--220~km\,s${^{-1}}$. On the contrary, after 150~kpc, $\sigma_{\rm p}$ is 
sharply increasing from a minimum of 200~km\,s${^{-1}}$, to reach 
400~km\,s$^{-1}$ at $R =$ 200~kpc, i.e. close to the cluster core 
(King) radius estimated  in the  visual Fornax Cluster Catalog  
(Ferguson \cite{ferguson89}).

\section{Ultra-compact objects and Fornax dwarf galaxies}
\label{sectUCDs}
In addition to unresolved GC candidates, we also targetted WFI slightly 
extended $V\le 20.5$ sources, of $FWHM$\,$\le$\,$1''$ but SExtractor 
(Bertin \& Arnouts \cite{bertin96}) {\it stellarity index}$<$0.3, candidates 
for the so-called ``ultra-compact dwarfs'' (UCDs, Phillipps et~al. 
\cite{phillipps01}; Drinkwater et~al. \cite{drinkwater03}; Mieske et~al. 
\cite{mieske04}). Eight (five class A, see Table~\ref{uc}) of the 27 compact 
objects we observed have FLAMES velocities compatible with Fornax membership.
Five of them were previously uncatalogued. These objects do appear resolved in 
our best (0\farcs7 seeing)  $I$-band WFI images and correspond to subarcsec
i.e. half-light radii $<$50 pc. It is worth mentioning that one of them would 
be located (at least) 200~kpc from the central NGC~1399; however this possible 
new UCD has not a well-determined velocity, as it lies on the limit between 
class B and class C objects, and needs further confirmation.

We also detect 19 (15 class A+B) ``normal'' dwarf galaxies in Fornax, mostly 
nucleated, 11 without previous velocity measurement (see Table~\ref{dw}). 
FLAMES Fornax dwarfs show 
a global velocity dispersion of 413$\pm$9~km\,s${^{-1}}$, in good agreement 
with the 429$\pm$41~km\,s$^{-1}$ value found by Drinkwater et~al. 
(\cite{drinkwater01}) for 55 dwarfs in the whole cluster. 

\section{Concluding remarks}
The origin of ICGCs may be drawn from two main scenarios: (1) ICGCs are a 
primordial population, formed at high redshift at the very beginning of the 
Fornax cluster building-up, without being associated to any given galaxy;
(2) ICGCs are GCs stripped from Fornax (possibly infalling dwarf) galaxies. 

Using an average specific frequency (defined as the number of GCs per unit 
$M_V=-15$ of galaxy luminosity) of $S_N\sim4$ for all giant galaxies in Fornax 
(Harris \cite{harris91}), but excluding the central NGC 1399, we would expect 
to find roughly 500 GCs within $-11<M_V<-10$. Within our spatial coverage, we 
should observe a maximum of about 30 such luminous GCs at radii larger than 
50~kpc, assuming that stripped GCs follow the spatial distribution of 
galaxies. We count 20 such bright ICGCs beyond 50~kpc, which would imply a 
66\% stripping efficiency if it was the only process at play. This may be a 
lower estimate, considering the low $S_N$ values found recently by Bassino 
et~al. (\cite{bassino06b}) for some early-types in our fields. Such a high 
stripping efficiency is unlikely and suggests that part of the ICGCs
are a primordial intracluster population. 

Nevertheless, stripping of GCs clearly appears ongoing in Fornax, as in the 
pass-by encounter of NGC~1399 and NGC~1404 (Bekki et~al.\ \cite{bekki03b}). 
The high-velocity NGC~1427A ($\varv = 2028$~km\,s$^{-1}$) is also 
probably infalling towards the centre, considering its ``blown-off'' 
morphology. Georgiev et~al. (\cite{georgiev06}) recently studied this LMC-like
galaxy. Using their identification chart, four of their photometrically 
selected candidates are FLAMES-confirmed cluster members; in the same studied 
area, we find another they did not consider, being just brighter than their 
magnitude cut at $V=21$. In such cases, an efficient stripping of GCs could 
naturally explain the high $S_N>10$ of NGC~1399 (Hilker et~al. 
\cite{hilker99}; but see Dirsch et~al. \cite{dirsch04} or Forte et~al. 
\cite{forte05}), at the center of the potential well.

Our observations show that the inner ICGC population ($R<150$~kpc) has
a velocity dispersion which decreases with radius, is likely relaxed
and is still probably associated to the (outer) halo of NGC~1399.
Beyond 150~kpc, a dynamically distinct, hotter population of GCs
appears, with a velocity dispersion which increases with radius  
until, at around 200~kpc, it reaches 
the dispersion of Fornax dwarf galaxies, thus 
suggesting a link between them. Drinkwater et~al. (\cite{drinkwater01}) 
proposed an infall scenario to explain a higher velocity dispersion of 
429~km\,s$^{-1}$ for Fornax dwarfs on a much larger scale, compared to 
308~km\,s$^{-1}$ for giant members.

The most massive (IC)GCs might also be regarded as extremely compact dwarfs
 rather than stellar clusters; the nature of so compact stellar systems 
still being debated, UCDs may be considered as ``transitional'' objects. 
Indeed, Ha\c{s}egan et~al. (\cite{hasegan05}) called them DGTOs for 
Dwarf-Globular Transition Objects. UCDs were first discovered in Fornax 
(Hilker et~al. \cite{hilker99}; Phillipps et~al. \cite{phillipps01}), but
their existence is now confirmed in Virgo (Jones et~al. \cite{jones06}) and 
they are probably present in more distant clusters like Abell 1689 
(Mieske et~al. \cite{mieske05}). It is not yet clear if UCDs are the remnant 
of dE(,N) tidally-truncated cores (Bekki et~al.\ \cite{bekki03a}) or were 
formed by aggregation of star clusters (Fellhauer \& Kroupa 
\cite{fellhauer02}). Another explanation could be that UCDs come simply from 
the direct evolution of very massive young clusters, as those formed in tidal 
tails of mergers (Kissler-Patig et~al. \cite{kissler06}). UCDs may then be 
downscaled versions of the tidal dwarf galaxies found in the simulations of 
Bournaud \& Duc (\cite{bournaud06}). Finally, Reverte et al. (\cite{reverte06})
have recently discovered two ultra-compact \ion{H}{ii}, strong starburst 
regions in H$\alpha$ images of two Abell clusters. Are they possible 
progenitors of UCDs?

Some authors tend to call ``UCDs'' all bright {\it unresolved} members in 
Fornax, so that they could rather be observing extremely bright but 
{\it genuine} GCs, like $\omega$ Cen in the Galaxy or Mayall II in M31 (Meylan 
et~al. \cite{meylan01}), which both exhibit some peculiarities but overall fit
well within the GC locus in Kormendy's (\cite{kormendy85}) diagrams. Here, we 
have reserved a conservative definition of ``ultra-compact'' only for the 
sources resolved in our 0$\farcs$7 images. The definitive distinction between 
UCDs and GCs, if any, may only be done through high-resolution HST imaging and 
spectroscopy of all compact bright Fornax members, to estimate their central 
velocity dispersion and infer their mass-to-light ratios. 

By means of wide-field VLT/FLAMES multi-object spectroscopy, we have confirmed 
149 bright GCs and 27 (including 8 ultra-compact) dwarf galaxies as likely 
members of the Fornax cluster; 134 of those GCs were previously uncatalogued. 
The superb radial velocity accuracy (10~km\,s${^{-1}}$, typically, up to 
$V\sim22$ magnitude targets) and multiplexity-spatial coverage of FLAMES in 
its GIRAFFE/MEDUSA mode allows us to give quantitative results on the 
kinematics in the Fornax central E--W strip, up to 220~kpc i.e., one cluster 
core radius off the central NGC~1399. Considering in particular the 61 GCs 
lying beetwen the Fornax galaxies, we find that their velocity dispersion 
-- after a decline from 350 to 200~km\,s$^{-1}$ at what could be considered 
the vanishing NGC~1399 halo, around 150~kpc -- next remarkably approaches 
400~km\,s${^{-1}}$ at $R \sim R_{\rm c} = 220$~kpc from it. This population of 
ICGCs could well not be relaxed yet.

It is tempting to compare the outermost ICGC dispersion to the value found
for an infalling population of Fornax dwarf galaxies 
(Drinkwater et~al. \cite{drinkwater01}). Other recent studies based on various 
techniques (distance determination via SBF, Dunn \& Jerjen \cite{dunn06}; 
X-ray gas properties: Scharf et al. \cite{sharf05}) suggest that 
accretion/collapse is still ongoing in Fornax which thus appears to be a 
rather dynamically young, compact cluster. More FLAMES observations of 
globular clusters in the outskirts of Fornax, including towards the Fornax A 
(NGC~1316) subgroup, may definitely confirm this trend for such intergalactic
globular clusters and provide new insights on the building-up of galaxy 
clusters. 

\begin{acknowledgements}
We thank the referee, M. Drinkwater, for useful comments.
We are indebted to 
Dominique Proust and Mathieu Puech for taking great care of the FLAMES
observations during the two nights of GEPI/Observatoire de Paris guaranteed 
time used for this work. 
We acknowledge Francine Leeuwin for leading the imaging part of 
the project in its early beginnings. GB is supported at the IAA/CSIC by an I3P 
contract (I3P-PC2005-F) funded by the European Social Fund, with additional 
support by DGI grant AYA 2005-07516-C02-01 and the Junta de Andaluc\'\i a.
This work made use of the HyperLEDA database ({\tt http://leda.univ-lyon1.fr}).
\end{acknowledgements}

\onltab{1}{
\begin{table*}
\renewcommand{\footnoterule}{}
\caption{Fornax class A and B globular clusters confirmed by the FLAMES study, identified by 
their WFI source number with their $J$2000.0 coordinates, heliocentric radial 
velocity $\varv$ and associated error $\delta\varv$ as estimated by 
{\tt fxcor}, as well as the resulting CCF fit $\mathcal{R}$ coefficient. 
The comments include a
comparison with GCs observed in previous studies: Mieske et~al. 
(\cite{mieske04}), prefix ``fcos'', and Dirsch et al. (\cite{dirsch04}),
prefix ``dir''. 
Among the brightest objects, gc302.3, gc46.30, gc241.1, gc85.30 and gc271.5 show 
marginal signs of resolution, with $FWHM$ about 10\% 
larger than the seeing. Higher resolution data would be needed to classify them as 
likely new UCDs.
}
\label{gc}
\centering
\begin{tabular}{l c c rl rl l}
\hline \hline
Id$_{\rm{WFI}}$ & $\alpha\ $($J$2000.0) & $\delta\ $($J$2000.0) &
$\varv$ (km\,s$^{-1}$) & $\delta\varv$ & $\mathcal{R}$ & Class & Comments \\
\hline
gc302.3            & 03:37:43.560 & -35:22:51.20 & 1419 & 11    & 6.8   & A &\\                                  
gc46.30            & 03:37:27.562 & -35:30:12.54 & 1913 & 5     & 13.9  & A & ICGC\\                                  
gc241.1  & 03:39:17.672 & -35:25:29.86 & 1027 & 8     & 10.9  & A & fcos1-060=980$\pm$45\\                               
gc76.40            & 03:36:57.252 & -35:29:56.77 & 1246 & 7     & 16.7  & A &\\                                 
gc230.7            & 03:36:34.335 & -35:19:32.58 & 1861 & 5     & 16.0  & A &\\                                 
gc85.30            & 03:37:45.066 & -35:29:01.28 & 1657 & 9     & 7.6   & A &\\                                 
gc235.7            & 03:36:12.687 & -35:19:11.57 & 1310 & 21    & 3.3   & B &\\                       
gc41.40            & 03:36:51.638 & -35:30:38.75 & 1468 & 17    & 4.9   & A &\\                            
gc89.40            & 03:36:55.337 & -35:29:37.71 & 1209 & 9     & 12.1  & A &\\                                 
gc271.5            & 03:37:05.708 & -35:37:32.09 & 1520 & 7     & 8.6   & A & ICGC\\                                 
gc131.6            & 03:40:32.511 & -35:36:23.04 & 1465 & 5     & 16.0  & A & ICGC\\                                 
gc199.6            & 03:40:22.907 & -35:33:50.74 & 1040 & 5     & 12.5  & A & ICGC\\                                 
gc6.300            & 03:40:21.409 & -35:24:27.64 & 1752 & 5     & 10.3  & A & ICGC\\                                 
gc129.2            & 03:41:29.535 & -35:19:48.60 & 1473 & 6     & 10.6  & A & ICGC\\                                 
gc56.40            & 03:37:03.991 & -35:30:16.48 & 1372 & 9     & 6.6   & A &\\                                 
gc323.6            & 03:35:21.541 & -35:14:42.23 & 1464 & 5     & 14.8  & A &\\                                 
gc144.6            & 03:35:38.864 & -35:21:53.40 & 1388 & 32    & 3.8   & B & ICGC\\           
gc35.30 & 03:37:42.235 & -35:30:33.85 & 1319 & 9     & 7.1   & A & fcos2-2094=1462$\pm$121  \\                               
gc44.40            & 03:37:00.037 & -35:30:36.19 & 1268 & 13    & 7.4   & A &\\                                 
gc613.2            & 03:38:37.978 & -35:23:32.57 & 1050 & 15    & 3.6   & B &\\                         
gc317.2 & 03:38:11.700 & -35:27:15.88 & 1434 & 8     & 6.6   & A & fcos2-2127=1476$\pm$201\\                                 
gc381.6 & 03:37:46.708 & -35:34:42.16 & 1386 & 13    & 4.2   & B & ICGC; fcos2-086=1400$\pm$50\\                           
gc248.7            & 03:36:32.793 & -35:18:30.28 & 1611 & 6     & 11.6  & A &\\                                 
gc152.1            & 03:42:00.125 & -35:19:32.49 & 947  & 15    & 3.2   & B &\\                        
gc359.8 & 03:39:19.063 & -35:34:06.66 & 1526 & 10    & 4.5   & A & fcos1-2050=1635$\pm$104\\                           
gc319.1   & 03:38:49.845 & -35:23:35.56 & 972  & 16    & 4.3   & A & dir76:63=933$\pm$29, fcos1-064=900$\pm$85\\                     
gc365.2            & 03:38:14.179 & -35:26:43.35 & 1143 & 19    & 4.6   & A &\\                           
gc324.8            & 03:39:06.040 & -35:34:49.52 & 1540 & 10    & 4.4   & A &\\                                 
gc220.8            & 03:36:50.357 & -35:20:16.82 & 1606 & 6     & 10.4  & A & ICGC\\                                 
gc69.20            & 03:41:05.011 & -35:22:08.55 & 1634 & 8     & 7.5   & A & ICGC\\                                 
gc357.2    & 03:38:38.135 & -35:26:46.39 & 1800 & 13    & 3.8   & B & dir77:2=1815$\pm$34\\                           
gc290.6            & 03:35:42.523 & -35:13:51.71 & 1901 & 16    & 3.9   & B & ICGC\\        
gc396.2   & 03:38:17.075 & -35:26:30.66 & 1235 & 12    & 6.0   & A & dir82:57=1232$\pm$22, fcos0-2089=1294$\pm$95\\                                    
gc69.70            & 03:36:01.104 & -35:25:43.07 & 1389 & 8     & 11.0  & A &\\                                 
gc154.7            & 03:38:26.616 & -35:41:42.96 & 1739 & 15    & 5.6   & A & ICGC\\                                 
gc216.7            & 03:38:30.776 & -35:39:56.63 & 1676 & 9     & 5.1   & A &\\                                  
gc410.7            & 03:38:30.401 & -35:34:20.49 & 1263 & 10    & 5.1   & A &\\                                 
gc70.50            & 03:40:02.811 & -35:38:57.16 & 1467 & 5     & 13.2  & A &\\                                 
gc175.1 & 03:39:05.005 & -35:26:53.47 & 1057 & 10 & 6.8 & A & 
fcos0-2092=970$\pm$205\\                                
gc346.6            & 03:35:21.028 & -35:13:53.22 & 1374 & 6     & 15.6  & A &\\                                 
gc212.5            & 03:39:49.163 & -35:34:46.17 & 1770 & 8     & 6.1   & A & ICGC\\                                 
gc272.7            & 03:36:27.090 & -35:17:33.30 & 1573 & 10    & 8.0   & A &\\                                 
gc465.7            & 03:38:43.527 & -35:33:07.70 & 1743 & 13    & 4.0   & A &\\                           
gc269.8            & 03:38:53.244 & -35:36:51.71 & 2000 & 13    & 4.7   & A &\\                           
gc376.8            & 03:39:15.834 & -35:34:55.97 & 1506 & 12    & 3.7   & B &\\                            
gc155.4            & 03:37:21.301 & -35:27:53.20 & 1218 & 12    & 5.8   & A & ICGC\\                                 
gc247.7            & 03:36:36.134 & -35:18:38.60 & 1447 & 8     & 8.9   & A &\\                                 
gc445.7            & 03:38:09.204 & -35:35:06.68 & 1773 & 22    & 4.0   & B &\\                            
gc21.70            & 03:35:59.574 & -35:26:56.58 & 1272 & 12    & 7.8   & A &\\                                 
gc260.7            & 03:36:30.136 & -35:17:54.08 & 1879 & 7     & 10.3  & A &\\                                 
gc378.8            & 03:39:09.177 & -35:34:57.94 & 1751 & 10    & 5.3   & A &\\                                 
gc555.2 & 03:38:30.730 & -35:24:40.15 & 1323 & 24    & 4.2   & A & dir77:40=1334$\pm$13\\                           
gc362.5            & 03:36:58.015 & -35:34:31.98 & 950  & 7     & 4.8   & B &\\                    
gc398.5            & 03:37:21.128 & -35:32:57.02 & 1713 & 11    & 5.5   & A & ICGC\\                                 
gc373.7   & 03:38:14.751 & -35:33:24.32 & 1347 & 11    & 5.1   & A & dir90:15=1298$\pm$31\\                        
gc120.6            & 03:40:44.611 & -35:36:46.74 & 1241 & 9     & 5.5   & A & ICGC\\                                 
gc156.2            & 03:38:13.538 & -35:28:55.89 & 1606 & 14    & 4.8   & A &\\                                 
gc311.6            & 03:37:59.444 & -35:36:09.18 & 1564 & 12    & 4.6   & A & ICGC\\                         
gc61.10            & 03:38:49.499 & -35:29:38.93 & 1949 & 12    & 3.6   & B &\\                     
gc18.70            & 03:36:31.257 & -35:26:58.22 & 1320 & 7     & 8.0   & A & ICGC\\                                 
gc107.4            & 03:36:55.732 & -35:29:21.55 & 1260 & 6     & 12.1  & A &\\                                 
gc387.2            & 03:38:23.279 & -35:26:32.72 & 1505 & 9     & 5.4   & A &\\                                 
gc428.7            & 03:41:20.356 & -35:28:46.30 & 1514 & 6     & 10.3  & A & ICGC\\                                 
gc350.2            & 03:41:20.539 & -35:12:53.66 & 1415 & 9     & 7.3   & A & ICGC\\                                 
gc388.5            & 03:36:58.337 & -35:32:06.84 & 1314 & 10    & 6.2   & A &\\                                 
gc382.5            & 03:40:06.761 & -35:29:27.32 & 1274 & 5     & 12.0  & A & ICGC\\                                 
gc153.8            & 03:36:48.830 & -35:22:46.37 & 1223 & 7     & 9.7   & A & ICGC\\                                 
gc172.2            & 03:38:30.172 & -35:28:47.81 & 1946 & 8     & 5.0   & A &\\                                  
gc115.3  & 03:38:04.436 & -35:28:11.22 & 1624 & 15    & 5.7   & A & dir91:109=1639$\pm$19\\                                 
gc133.3            & 03:37:47.017 & -35:27:47.96 & 1835 & 20    & 4.2   & B &\\                           
gc401.6            & 03:37:33.894 & -35:32:46.91 & 1748 & 8     & 6.1   & A & ICGC\\                                 
gc159.5            & 03:40:02.534 & -35:36:48.68 & 1741 & 6     & 11.3  & A &\\                                 
gc101.1            & 03:39:22.021 & -35:28:44.79 & 693  & 7     & 4.3   & B & ICGC\\                           
gc225.6            & 03:37:46.793 & -35:39:23.54 & 1339 & 16    & 3.8   & B & ICGC\\                           
gc280.1            & 03:39:03.854 & -35:24:28.39 & 1880 & 16    & 3.6   & B &\\                           
gc77.20            & 03:41:21.179 & -35:21:46.23 & 2076 & 20    & 3.5   & B & ICGC\\                   
gc236.6            & 03:40:49.601 & -35:32:46.80 & 1479 & 5     & 13.0  & A & ICGC\\                                 
gc114.7            & 03:36:10.836 & -35:24:22.46 & 1362 & 6     & 8.1   & A &\\                                 
gc152.5            & 03:39:39.420 & -35:36:58.51 & 1729 & 7     & 8.4   & A & ICGC\\                                 
gc223.5            & 03:40:00.990 & -35:34:35.59 & 1387 & 7     & 11.0  & A & ICGC\\                                 
gc12.70            & 03:36:12.131 & -35:27:07.71 & 1419 & 6     & 8.9   & A &\\                                 
gc175.6            & 03:40:38.831 & -35:34:43.12 & 1693 & 7     & 9.4   & A & ICGC\\                                 
gc300.6            & 03:37:59.568 & -35:36:25.14 & 1949 & 9     & 4.8   & B & ICGC\\                           
gc387.8            & 03:38:58.137 & -35:35:24.77 & 785  & 9     & 6.2   & A &\\                                 
gc19.40            & 03:37:00.336 & -35:31:13.97 & 1265 & 7     & 10.0  & A &\\                                 
gc289.7            & 03:38:46.424 & -35:37:23.04 & 2052 & 10    & 5.6   & A &\\                           
gc466.7            & 03:38:08.804 & -35:32:25.53 & 1801 & 10    & 3.6   & B &\\                     
gc91.20            & 03:38:41.920 & -35:29:48.47 & 1484 & 9     & 3.8   & B &\\                             
gc43.40            & 03:37:13.122 & -35:30:41.35 & 1465 & 11    & 4.0   & B &\\                             
gc221.2            & 03:38:37.212 & -35:28:12.73 & 1578 & 15    & 5.5   & A &\\                            
gc7.700            & 03:36:03.886 & -35:27:26.42 & 1411 & 11    & 6.7   & A &\\                                 
gc280.7            & 03:38:13.399 & -35:37:37.88 & 981  & 13    & 3.9   & B & ICGC\\                           
gc414.7         & 03:38:59.310 & -35:33:43.43 & 2016 & 15    & 3.7   & B & dir84:11=2059$\pm$28\\                     
gc102.2  & 03:38:16.653 & -35:29:34.94 & 1669 & 9     & 4.6   & B & dir89:74=1664$\pm$28 \\                         
gc71.60            & 03:40:23.320 & -35:38:32.02 & 1301 & 8     & 8.8   & A &\\                                 
gc89.10            & 03:39:05.613 & -35:28:59.31 & 1037 & 12    & 4.0   & B &\\                           
gc212.2            & 03:38:28.441 & -35:28:20.97 & 1804 & 19    & 3.9   & B &\\                           
gc311.3            & 03:40:45.968 & -35:14:51.64 & 1632 & 17    & 4.5   & A & ICGC\\                                 
gc177.6            & 03:37:26.253 & -35:41:05.69 & 985  & 8     & 5.8   & A & ICGC\\                                 
gc90.40            & 03:36:59.578 & -35:29:39.25 & 1304 & 7     & 7.6   & A &\\                                 
gc163.7            & 03:41:10.461 & -35:34:57.47 & 1448 & 7     & 8.4   & A & ICGC\\                                 
gc417.5            & 03:39:39.759 & -35:25:51.69 & 1141 & 15    & 4.5   & A & ICGC\\                                 
gc170.7            & 03:36:36.316 & -35:21:58.68 & 1472 & 10    & 7.9   & A & ICGC\\                                 
gc395.7            & 03:41:03.826 & -35:26:34.05 & 1459 & 8     & 6.3   & A & ICGC\\                                 
gc115.4            & 03:40:12.554 & -35:21:11.72 & 1233 & 8     & 6.8   & A & ICGC\\                                 
gc2.100            & 03:42:11.924 & -35:24:42.93 & 1413 & 9     & 5.5   & A &\\                                 
gc269.5            & 03:39:43.107 & -35:33:10.41 & 1274 & 12    & 3.8   & B & ICGC\\                         
gc4.700            & 03:36:06.204 & -35:27:32.82 & 1252 & 9     & 4.8   & A &\\                   
gc172.4            & 03:40:11.404 & -35:19:29.25 & 1365 & 9     & 5.2   & A & ICGC\\                                 
gc345.7  & 03:38:13.066 & -35:33:52.43 & 1566 & 10    & 3.4   & B & dir90:9=1523$\pm$29\\                 
gc70.70            & 03:36:09.122 & -35:25:43.69 & 1403 & 8     & 5.9    & B&\\                                  
gc73.10            & 03:42:13.711 & -35:22:41.10 & 1398 & 12    & 5.1    & B&\\                     
gc302.6            & 03:35:50.492 & -35:15:24.22 & 1166 & 6     & 8.0   & A & ICGC\\                                 
gc332.7            & 03:41:13.612 & -35:29:28.90 & 1527 & 13    & 4.8   & A & ICGC\\                                 
gc30.60            & 03:35:21.876 & -35:14:05.41 & 1249 & 16    & 4.4    & B& \\                     
gc381.7            & 03:41:05.903 & -35:26:38.00 & 607  & 9     & 6.2   & A & ICGC\\                                 
gc172.2            & 03:41:13.713 & -35:18:17.44 & 1861 & 9     & 3.9    & B& ICGC\\                 
gc317.5            & 03:39:39.846 & -35:31:53.61 & 1689 & 11    & 7.3   & A & ICGC\\                                 
gc201.1            & 03:41:48.413 & -35:17:40.25 & 1295 & 10    & 6.7   & A & ICGC\\                                 
gc304.8            & 03:41:49.465 & -35:30:12.68 & 1570 & 12    & 3.1    & B& ICGC\\                 
gc32.10            & 03:42:16.258 & -35:24:06.13 & 1399 & 12    & 5.6   & A &\\                                 
gc13.40            & 03:40:08.114 & -35:24:18.05 & 1532 & 8     & 8.6   & A & ICGC\\                                 
gc173.7            & 03:36:23.575 & -35:21:52.32 & 1413 & 10    & 6.0   & A & ICGC\\                                 
gc459.5            & 03:40:12.978 & -35:27:03.48 & 1366 & 9     & 7.2   & A & ICGC\\                           
gc57.40            & 03:36:53.187 & -35:30:14.43 & 1342 & 16    & 6.9   & A &\\                                 
gc85.10            & 03:42:17.040 & -35:22:08.41 & 1578 & 11    & 7.0   & A &\\                                 
gc164.6            & 03:40:24.712 & -35:35:13.69 & 949  & 10    & 5.6   & A & ICGC\\                                 
gc173.5            & 03:40:09.731 & -35:36:11.17 & 1898 & 9     & 5.2   & A &\\                                 
gc39.70            & 03:36:01.107 & -35:26:22.28 & 1324 & 19    & 5.9   & A & Asymetric CCF\\                         
gc516.5            & 03:39:40.293 & -35:28:53.71 & 1414 & 14    & 4.1    & B& ICGC\\                     
gc6.400            & 03:39:57.562 & -35:24:33.00 & 1564 & 9     & 4.9   & A & ICGC\\                           
gc67.10            & 03:42:14.505 & -35:23:00.09 & 1708 & 12    & 3.9    & B&\\      
gc325.6            & 03:35:18.844 & -35:12:51.13 & 1429 & 13    & 4.5   & A &\\               
gc391.5            & 03:39:44.276 & -35:29:15.90 & 1520 & 10    & 6.8   & A & ICGC\\                                 
gc441.5            & 03:39:59.951 & -35:26:31.57 & 1775 & 9     & 4.5   & A & ICGC\\                           
gc388.2            & 03:41:11.311 & -35:09:20.21 & 555  & 12    & 4.0    & B& ICGC (fcc160 GC?)\\          
gc398.3            & 03:40:40.878 & -35:12:30.49 & 1250 & 11    & 3.4    & B& ICGC\\  
gc74.10            & 03:42:18.711 & -35:22:40.23 & 1468 & 13    & 4.7   & A &\\                                  
gc375.1            & 03:41:51.519 & -35:11:24.91 & 1907 & 27    & 2.9    & B& ICGC\\    
gc467.5            & 03:40:11.597 & -35:27:09.52 & 1424 & 9     & 6.0   & A & ICGC\\                                 
gc187.2            & 03:41:05.002 & -35:17:45.30 & 1575 & 12    & 4.1    & B& ICGC\\                 
gc1387se           & 03:36:58.701 & -35:30:36.21 & 1273 & 7     & 17.9  & A& \\                                 
gc1387sw           & 03:36:55.132 & -35:30:35.93 & 1340 & 10    & 11.5  & A& \\                                 
gc1404e            & 03:38:54.587 & -35:35:30.18 & 1911 & 44    & 4.7   & A& \\                            
gc1404n            & 03:38:52.046 & -35:35:14.81 & 1816 & 16    & 9.1   & A& \\                                 
gc1404s            & 03:38:51.609 & -35:36:10.49 & 2174 & 16    & 4.3    & B&\\                            
gc1404w            & 03:38:49.016 & -35:35:33.05 & 1730 & 36    & 3.4    & B&\\                            
gc1375s            & 03:35:17.214 & -35:16:07.65 & 785  & 12    & 8.5   & A & Broad CCF\\                        
gc1375e            & 03:35:18.768 & -35:15:53.06 & 852  & 6     & 14.5  & A &\\ 
\hline
\end{tabular}
\end{table*} 
}
\onltab{2}{
\begin{table*}
\caption{Fornax class A and B ultra-compact objects confirmed by the FLAMES 
study, identified by 
their WFI source number with their $J$2000.0 coordinates, heliocentric radial 
velocity $\varv$ and associated error $\delta\varv$ as estimated by {\tt fxcor},
as well as the resulting CCF fit $\mathcal{R}$ coefficient. 
The comments include a comparison with objects observed in previous studies: Phillipps et~al. 
(\cite{phillipps01}), prefix ``2dF''; Mieske et~al. 
(\cite{mieske04}), prefix ``fcos''; and Dirsch et~al. (\cite{dirsch04}),
prefix ``dir''. 
}
\label{uc}
\centering
\begin{tabular}{l c c rl rl l}
\hline \hline
Id$_{\rm{WFI}}$ & $\alpha\ $($J$2000.0) & $\delta\ $($J$2000.0) &
$\varv$ (km\,s$^{-1}$)& $\delta\varv$ & $\mathcal{R}$ & Class & Comments \\
\hline
ucdB	& 03:39:35.920	& -35:28:24.59	& 1878& 5	& 15.8	& A & fcos1-2083=1848$\pm$85, 2dF=1920$\pm$40\\
uc218.7	& 03:38:23.416	& -35:39:53.32	& 757 & 21	& 3.8	& B & \\
uc257.5	& 03:37:03.233	& -35:38:04.37	& 1551& 5	& 18.6	& A & fcos2-2031=1571$\pm$75, 2dF=1491$\pm$39\\
uc329.7	& 03:36:27.711	& -35:14:13.91	& 1386& 4	& 29.6	& A & \\
uc411.2	& 03:41:32.855	& -35:11:37.30	& 2076& 11	& 3.3	& B &\\
uc464.5	& 03:39:43.521	& -35:26:59.25	& 1274& 7	& 9.3	& A & \\
uc82.40	& 03:36:58.524	& -35:29:45.37	& 1379& 19	& 5.2	& B & In NGC 1387?\\
ucdA	& 03:38:06.298	& -35:28:58.49	& 1234&	5	&17.3	& A & fcos2.2111=1280$\pm$58, dir91:93=1247$\pm$16, dir90:86=1218$\pm$9\\
\hline
\end{tabular}
\end{table*} 
}
\onltab{3}{
\begin{table*}
\caption{Fornax FLAMES dwarves (all classes). 
Names are from Ferguson (\cite{ferguson89}) or ``newdw'' for the newly detected dwarves, 
with the $J$2000.0 centre coordinates, heliocentric radial 
velocity $\varv$ and associated error $\delta\varv$ as estimated by {\tt fxcor},
as well as the resulting CCF fit $\mathcal{R}$ coefficient. 
The comments include a comparison with velocities available in the literature, when available. 
}
\label{dw}
\centering
\begin{tabular}{l c c rl rl l}
\hline \hline
Name & $\alpha\ $($J$2000.0) & $\delta\ $($J$2000.0) &
$\varv$ (km\,s$^{-1}$) & $\delta\varv$ & $\mathcal{R}$ & Class & Comments \\
\hline
fcc208     & 03:38:18.791 & -35:31:50.69  & 1756 & 9     & 5.8   & A & 1720$\pm$50 \\                   
fcc194     & 03:37:17.933 & -35:41:57.30  & 1295 & 15    & 4.3   & A & 1237$\pm$84   \\                       
newdw1     & 03:41:18.114 & -35:28:37.46  & 1750 & 15    & 4.0   & B & Previously uncatalogued  \\          
fcc259     & 03:41:07.230 & -35:30:52.10  & 1770 & 17    & 3.5   & B & 1st velocity determination \\              
fcc227     & 03:39:50.085 & -35:31:22.27  & 700  & 50    & 2.5   & C & 1st velocity determination, very uncertain   \\                   
fcc241     & 03:40:23.535 & -35:16:35.86  & 1824 & 7     & 10.2  & A & 2045$\pm$107 \\                  
fcc1554    & 03:41:59.442 & -35:20:56.18  & 1612 & 4     & 21.8  & A & 1642$\pm$52       \\                  
newdw2     & 03:39:56.441 & -35:37:20.83  & 1770 & 14    & 4.0   & C & Previously uncatalogued        \\                
fcc266     & 03:41:41.327 & -35:10:13.26  & 1555 & 3     & 30.3  & A & 1551$\pm$39         \\               
fcc160     & 03:41:12.880 & -35:09:31.48  & 548  & 8     & 7.9   & A & 1493$\pm$59           \\             
fcc272     & 03:42:11.300 & -35:26:35.00  & 1608 & 15    & 4.5   & B & 1st velocity determination       \\               
fcc160     & 03:36:04.050 & -35:23:20.00  & 1955 & 8     & 7.1   & A & 1st velocity determination         \\                 
fcc171     & 03:36:37.270 & -35:23:09.25  & 1395 & 5     & 12.0  & A & 1st velocity determination           \\               
newdw3     & 03:36:59.800 & -35:20:36.00  & 1964 & 30    & 3.6   & B & Previously uncatalogued         \\
newdw4     & 03:35:35.670 & -35:17:26.50  & 2373 & 14    & 3.7   & C & Previously uncatalogued           \\       
fcc182     & 03:36:54.303 & -35:22:28.78  & 1693 & 3     & 43.4  & A & 1669$\pm$11             \\           
fcc156     & 03:35:42.750 & -35:20:18.80  & 1405 & 22    & 2.8   & C & 1st velocity determination   \\
fcc215     & 03:38:37.655 & -35:45:27.39  & 1964 & 20    & 3.7   & B & 1st velocity determination     \\      
fcc222     & 03:39:13.319 & -35:22:16.74  & 800  & 5     & 12.8  & A & 792$\pm$26                \\        
\hline
\end{tabular}
\end{table*} 
}
\end{document}